\documentclass[twocolumn,aps,prb,superscriptaddress,floatfix]{revtex4-1}
\usepackage{hyperref}
\usepackage{color}
\usepackage{amsmath}
\usepackage{amssymb}
\usepackage{graphicx}
\usepackage{esint}

\newcommand{\Ztwo}{$\mathbb{Z}_2$}

\newcommand{\SUtwo}{SU(2)}
\newcommand{\Uone}{U(1)}

\graphicspath{{./}{./fig/}}

\begin{document}
\title{
  Random-bond disorder in two-dimensional non-collinear \emph{XY} antiferromagnets:\\
From quasi-long-range order to spin glass
}

\author{Santanu Dey}
\affiliation{Institut f\"ur Theoretische Physik and W\"urzburg-Dresden Cluster of Excellence ct.qmat, Technische Universit\"at Dresden,
01062 Dresden, Germany}
\author{Eric C. Andrade}
\affiliation{Instituto de F\'{i}sica de S\~ao Carlos, Universidade de S\~ao Paulo, C.P. 369,
S\~ao Carlos, SP,  13560-970, Brazil}
\author{Matthias Vojta}
\affiliation{Institut f\"ur Theoretische Physik and W\"urzburg-Dresden Cluster of Excellence ct.qmat, Technische Universit\"at Dresden,
01062 Dresden, Germany}

\begin{abstract}
We study effects of quenched bond disorder in frustrated easy-plane antiferromagnets in two space dimensions, using a combination of analytical and numerical techniques.
We consider local-moment systems which display non-collinear long-range order at zero
temperature, with the antiferromagnetic triangular-lattice \emph{XY} model as a prime example.
We show that (i) weak bond disorder transforms the clean ground state into a state with quasi-long-range order, and (ii) strong bond disorder results in a short-range-ordered
glassy ground state.
Using extensive Monte Carlo simulations, we also track the fate of the quasi-long-range-ordered phase at finite temperature and the associated Kosterlitz-Thouless-like transition. Making contact with previous studies of disordered \emph{XY} models, we discuss similarities and differences concerning the interplay of disorder and frustration.
Our results are also of relevance to non-collinear Heisenberg magnets in an external magnetic field.
\end{abstract}

\date{\today}
\maketitle


\section{Introduction}

The interplay of quenched disorder and topology is a multi-facetted and timely topic in condensed mater physics. On the one hand, gapped topological states of matter are generically stable against weak randomness; for symmetry-protected topological states this applies only if the protecting symmetries are preserved by the disorder.\cite{qizhang11} On the other hand, certain types of disorder may induce new effective degrees of freedom or nucleate topological defects, such that even nominally weak disorder qualitatively changes the system's properties.\cite{willans10,senmoessner15,kimchi17}

For easy-plane magnets in two space dimensions, with a local {\Uone} (or \emph{XY}) degree of freedom, the finite-temperature behavior is governed by vortices which are stable topological defects with an integer winding number; this also applies to arrays of Josephson junctions between superconducting islands and similar settings where the local degree of freedom represents a condensate phase. While long-range order (LRO) is realized at temperature $T=0$, the low-temperature phase displays quasi-long-range order (QLRO) which gives way to a thermally disordered high-temperature phase via a Kosterlitz-Thouless (KT) vortex unbinding transition.\cite{kosterlitz73}
For frustrated easy-plane magnets, the behavior can be more complicated, as chirality becomes a relevant degree of freedom.\cite{kawamura98} 
For the triangular-lattice \emph{XY} model it has been argued that two thermal phase transitions occur, a low-$T$ spin transition -
 which was assumed to be of KT type but whose nature has not been clarified beyond doubt - where QLRO is lost and a higher-$T$ Ising 
 transition where chirality order is lost \cite{olsson95,ozekiito03,obuchi12}. Equivalent physics occurs in fully frustrated \emph{XY} models
 \cite{hasenbusch05,hasenbusch05b,okumura11}.

The effect of quenched disorder has mainly been studied in the (unfrustrated) superconductor context: While random couplings do not change the physics qualitatively, random phases do: At $T=0$, LRO is destroyed in favor QLRO by infinitesimally small phase disorder, while large disorder leads to short-range-ordered state \cite{rubinstein83,nattermann95,cha95,scheidl96,tang96,koster97,carpentier98,mudry99,carpentier00,maucourt97,alba09,ozeki14}.
In contrast, the effect of randomness in the context of frustrated easy-plane magnets has not been studied in much detail. This is a particularly interesting topic given that recent work has shown that bond disorder in two-dimensional non-collinear magnets is able to destroy $T=0$ LRO via effective dipolar random fields \cite{dey19}.

In this paper, we consider the effect of bond-disorder in non-collinear easy-plane magnets in two space dimensions, using a combination of analytical arguments and numerical simulations, the latter performed for the classical triangular-lattice \emph{XXZ} antiferromagnet.
As for Heisenberg symmetry \cite{syrom15,dey19}, a single bond defect induces a dipolar spin texture, and a finite concentration of such defects destroys $T=0$ LRO. However, we show that in the easy-plane case the resulting state displays QLRO, in contrast to the Heisenberg case where it is a short-range-ordered spin glass. Such a spin glass is realized in the easy-plane case only beyond a critical level of bond disorder. The finite-disorder transition between the QLRO and glassy states most likely proceeds in two steps, with separate magnetic and chirality transitions, however, distinguishing the two transitions beyond doubt is beyond our numerical resolution. We also construct the finite-temperature phase diagram of the triangular-lattice \emph{XY} antiferromagnet, tracking the finite-temperature transition(s) as function of disorder.
The overall phenomenology which we find for bond-disordered non-collinear \emph{XY} models is not unlike that of random-phase \emph{XY} models, which can be rationalized stating that in both cases a single defect has a linear gradient coupling to the single Goldstone mode of the system. We point out that the physics of random-phase {\em fully frustrated} \emph{XY} models is very different: Here, phase disorder acts as a random field for chirality, thereby destroying any QLRO phase \cite{gupta99}.
We discuss applications of our results to compounds with frustrated easy-plane magnetism, and we point out connections to canted Heisenberg antiferromagnets in applied magnetic fields.

The remainder of the paper is organized as follows:
In Sec.~\ref{sec:models} we introduce the lattice models for frustrated easy-plane magnets and review the physics of isolated bond defects. We compare this to the behavior of random-phase \emph{XY} models, pointing out similarities and differences.
Sec.~\ref{sec:rg} sketches the field-theoretic treatment of dipolar disorder, highlighting differences between the cases of underlying \emph{XY} and Heisenberg symmetry.
In Sec.~\ref{sec:num} we present the results of large-scale numerical simulations for the classical \emph{XY} antiferromagnet on the triangular lattice, which we use to construct a quantitative phase diagram as function of temperature and disorder.
A concluding discussion closes the paper.


\section{Easy-plane magnets, \emph{XY} models, and disorder}
\label{sec:models}

In this section, we introduce two-dimensional lattice models for both non-collinear easy-plane magnetism and frustrated arrays of Josephson junctions and discuss similarities and differences upon including quenched disorder.

\subsection{Triangular-lattice \emph{XXZ} antiferromagnet}

In the context of Mott-insulating antiferromagnets, easy-plane magnetism is most naturally described by a spin-anisotropic \emph{XXZ} model
\begin{align}
\mathcal{H}_{\rm \emph{XXZ}} &=
        \sum_{\langle ij\rangle}
        J_{ij}
        \left(S^x_iS^x_j+S^y_iS^y_j
        + \lambda S^z_iS^z_j\right)
\label{eq:xxz}
\end{align}
of spins $S$ on a regular lattice of sites $i$, and ${\langle ij\rangle}$ denotes a summation over pairs of sites. The magnetic exchange coupling is $J>0$, its anisotropy is parameterized by $\lambda$ with $0\leq\lambda<1$ describing easy-plane magnetism with underlying {\Uone} symmetry. $\lambda=0$ corresponds to the pure \emph{XY} case, whereas $\lambda=1$ is the Heisenberg case with enhanced {\SUtwo} symmetry. A single-ion anisotropy has been omitted in $\mathcal{H}$; it will not change any qualitative conclusions.

Frustration can be induced in multiple ways and often results in non-collinear ground-state order. For concreteness, we will focus on the triangular lattice with nearest-neighbor interactions, $J_{ij}\equiv J$ for nearest neighbors and zero otherwise. For $\lambda<1$, the lowest-energy state displays three-sublattice $120^\circ$ order in the $xy$ plane. This state is chiral, i.e., the vector chirality
\begin{align}
        \vec{\kappa}_P &=
        \frac{2}{3\sqrt{3}}
        \sum_{ij\in P,j>i}
        \vec{S}_i\times\vec{S}_j
\label{eq:chir}
\end{align}
on each elementary plaquette $P$ displays Ising order along the $z$ axis, with the two directions corresponding to the handedness of the spiral.\cite{yosefin85,kawamura98} The spontaneously broken {\Uone} symmetry induces one Goldstone mode.

For the Heisenberg case, $\lambda=1$, chiral $120^\circ$ order is realized as well. However, the vector chirality can point along any direction in spin space, and the higher (broken) symmetry leads to two inequivalent types of Goldstone modes, corresponding to in-plane and out-of-plane rotations of the order, respectively. For $\lambda\lesssim 1$, the out-of-plane modes develop a gap scaling as $J\sqrt{1-\lambda}$.

\subsection{Random dipoles from bond disorder}

\begin{figure}
\includegraphics[width=0.95\columnwidth]{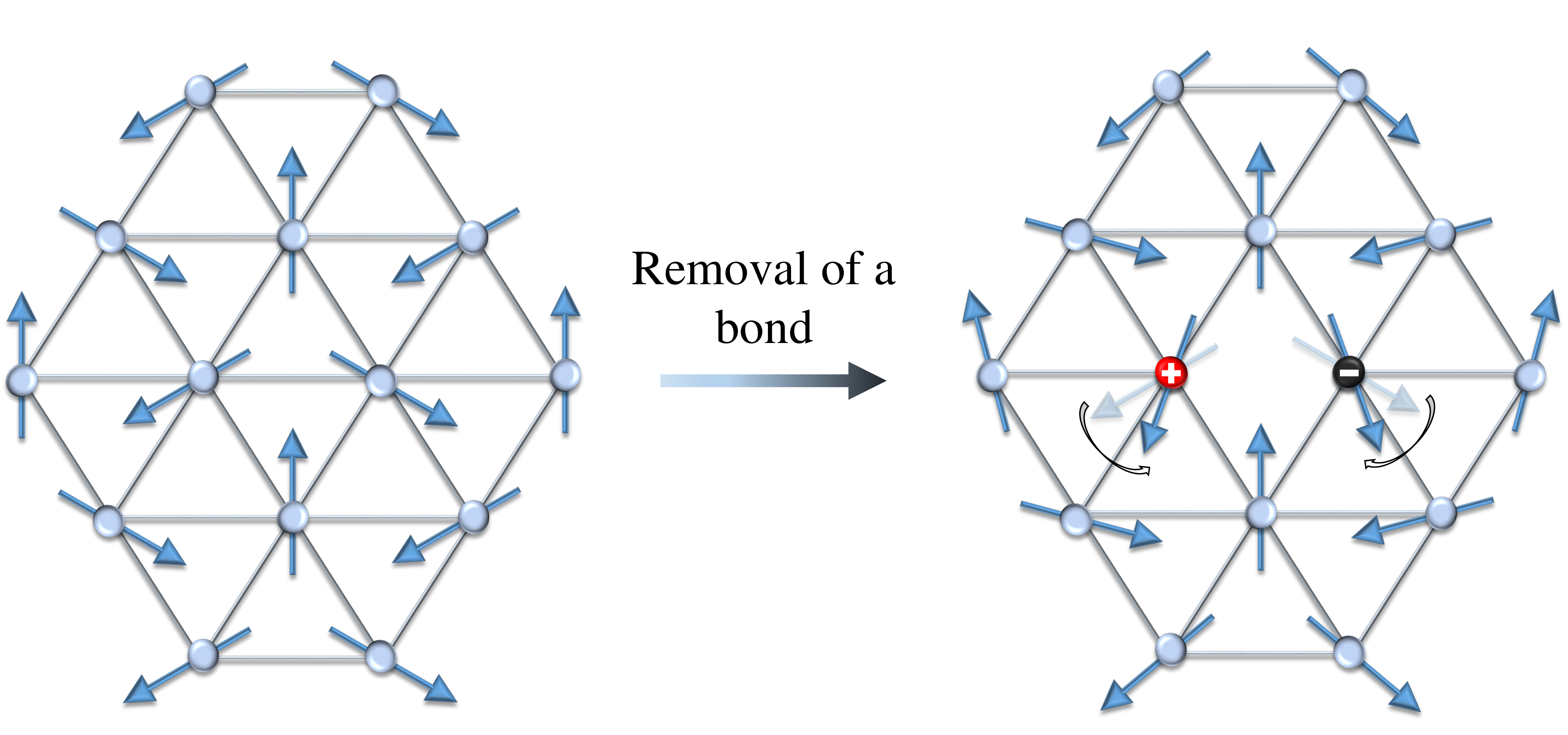}
\caption{
Dipolar distortion of the $120^\circ$ antiferromagnetic ground state of the triangular-lattice \emph{XY} model caused by a bond defect.
}
\label{fig:dipole}
\end{figure}

In a frustrated spin system, a single defect quite generically leads to a \emph{local} reduction of frustration, as it modifies the energy balance in the overconstrained system. This reduction of frustration causes a change of spin directions already at the classical level, which in turn induces a spin texture surrounding the defect. The symmetry of the texture depends on the geometry of both the lattice and the defect.

For a bond defect in a non-collinear chiral antiferromagnet with underlying continuous symmetry, it has been shown that the texture is dipolar,\cite{syrom15,dey19} Fig.~\ref{fig:dipole}: In response to the defect, the spins rotate in their ordering plane by an angle $\delta\varphi$ which depends on the distance $\vec r$ to the defect as
\begin{align}
\delta\varphi(\vec{r})
        =\kappa \delta J \frac{N_0^2}{\rho}
        \frac{\hat{e}\cdot\vec{r}}{r^2}
\label{eq:dipole}
\end{align}
where $N_0$ and $\rho$ are the ordered moment and the stiffness (per unit-cell area) of the background state, $\hat{e}$ is the oriented vector of the defect bond, and $\kappa$ is a numerical prefactor dependent on microscopic details [$\kappa=\sqrt{3}/(4\pi)$ for $120^\circ$ magnetic order on the triangular lattice]. The power-law decay in Eq.~\eqref{eq:dipole} arises from the coupling of the bond defect to the in-plane Goldstone mode of the bulk order.

A finite amount of bond disorder, e.g., a random distribution of nearest-neighbor $J_{ij}$, is then equivalent to a collection of random dipoles. It has been shown in Ref.~\onlinecite{dey19} that this destroy LRO in favor of a glassy state for non-collinear Heisenberg systems; in the present paper the focus is on the easy-plane case.

These random dipoles can be contrasted, e.g., to the cases of a vacancy in the triangular-lattice $120^\circ$ state which induces an octupolar texture,\cite{wollny11} and a vacancy in the stripy state of the square-lattice $J_1$-$J_2$ model which leads to a quadrupolar texture.\cite{weber12}


\subsection{Frustrated \emph{XY} models and random-phase disorder}

The low-energy degrees of freedom of easy-plane magnets are compact angular variables $\varphi_i$, subject to an \emph{XY}-type interaction. Such models are familiar in the description of phase degrees of freedom in arrays of Josephson-coupled superconducting grains or islands. In the latter context, couplings are usually ferromagnetic, and frustration can be introduced by bond-dependent phase shifts, leading to
\begin{align}
\mathcal{H}_{\rm \emph{XY}} = -\sum_{\langle ij\rangle} J_{ij}
        \cos\left(\varphi_i-\varphi_j-A_{ij} \right)\,.
\label{eq:ffxy}
\end{align}
Here $A_{ij}$ are phase shifts, which can arise from the magnetic flux of a field applied transverse to the plane of the Josephson array. While $A_{ij}=0$ yields an unfrustrated ferromagnetic ground state, maximal frustration is introduced by half a flux quantum per plaquette $P$, $\sum_{ij\in P} A_{ij}= \pi$ (mod $2\pi$). On the triangular lattice, this fully frustrated \emph{XY} model can be shown by a gauge transformation to be equivalent to the antiferromagnetic \emph{XY} model in Eq.~\eqref{eq:xxz} with $\lambda=0$, implying that its ground state spontaneously breaks both a {\Uone} and a {\Ztwo} symmetry, the latter corresponding to chirality.\cite{kawamura98}

Bond disorder can be introduced into the \emph{XY} model \eqref{eq:ffxy} by spatial variations of either the couplings $J_{ij}$ or the phase shifts $A_{ij}$. While random couplings have the same effect as in the easy-plane spin model, random phases are more subtle because $A_{ij}$ is a \emph{directed} quantity. This is particularly relevant for the fully frustrated case: Here, randomness in $J_{ij}$ does not couple to chirality and hence leaves the corresponding {\Ztwo} degeneracy of the ground state intact, whereas randomness in $A_{ij}$ couples to chirality and hence acts as a random field for this degree of freedom.
Remarkably, on the triangular lattice, a single defect of either type produces 
a dipolar texture in the
ground state akin to Eq.~\eqref{eq:dipole}. However, the sign of the dipolar distortion, 
encoded by the sign of $\hat{e}$,
behaves differently under the reversal of the global ground state's chirality 
in the two cases. For a single
$J_{ij}$ bond defect,  flipping the sign of the background chirality flips the 
sign of $\hat{e}$, thus reversing the 
polarity of the distortion texture. In contrast, for an $A_{ij}$ phase 
defect, the sign of $\hat{e}$ is always
fixed by the direction of $A_{ij}$ .

The effect of a finite amount of phase disorder has been studied in some detail: Extensive work on random-phase \emph{XY} (RPXY) models\cite{rubinstein83,nattermann95,cha95,scheidl96,tang96,koster97,carpentier98, mudry99,carpentier00,maucourt97,alba09,ozeki14} -- without frustration -- has established \cite{nattermann95,maucourt97} that the ground state has QLRO with power-law correlations at small disorder; this state continuously connects to the QLRO phase at finite $T$.
Upon increasing the disorder strength a KT-like transition destroys the QLRO phase in favor of a disordered phase with exponentially decaying correlations. The critical disorder strength decreases with increasing temperature. While it was initially argued from perturbative considerations \cite{rubinstein83} that a re-entrant behavior might be observed where the dependence of critical disorder strength for the KT transition is non-monotonic with temperature and at zero temperature the critical disorder strength vanishes, non-perturbative treatments \cite{nattermann95,scheidl96,tang96,carpentier98,carpentier00} and Monte-Carlo (MC) simulations \cite{maucourt97,alba09,ozeki14} have shown that the RPXY model does not exhibit any re-entrant behavior and the QLRO phase survives upto a finite critical value of disorder strength even at $T=0$.

Extensions of the RPXY model to include frustration were considered later in the context of disordered Josephson-junction arrays. For the fully frustrated case it has been shown,\cite{gupta99} using an argument along the lines of Imry and Ma,\cite{imry75} that the random-field nature of the random phases destroys the chiral order of the clean case already for infinitesimal disorder. The resulting phase is disordered with a finite correlation length even at $T=0$.
We recall that the physics of the random-bond easy-plane antiferromagnet is expected to be different because randomness in the couplings does \emph{not} linearly couple to chirality.


\section{Analytical considerations}
\label{sec:rg}

In this section, we give a few analytical arguments for the behavior of bond-disordered non-collinear easy-plane antiferromagnets, based on parallels to RPXY models. We also discuss the key difference between the easy-plane and Heisenberg cases.


\subsection{Quasi-long-range order from random dipoles}

In the limit of weak disorder, it is sufficient to analyze small fluctuations about the ordered state. These are governed by the interplay of the Goldstone mode and the dipolar disorder from bond defects. These ingredients are identical to that of the RPXY model -- without frustration -- in the same limit of weak disorder: For this model \eqref{eq:ffxy}, a single phase defect introduces a dipolar texture in the otherwise ferromagnetic ground state of the $\varphi_i$.

Therefore we can adopt knowledge gathered for the RPXY model in two space dimensions.\cite{rubinstein83,nattermann95,scheidl96,tang96} As a result of spin-wave fluctuations, the disorder-averaged phase--phase correlation function decays in a power-law fashion
\begin{equation}
[\exp(i\varphi_i-i\varphi_j)]_{\rm avg} \propto \frac{1}{|\vec{r}_i-\vec{r}_j|^\eta}
\end{equation}
where the decay exponent obeys
\begin{equation}
\label{eq:eta}
\eta =\frac{1}{2\pi} \left(\frac{T}{J}+\Delta^2\right)
\end{equation}
and $\Delta^2$ parameterizes the variance of the (Gaussian) distribution of random phases. For the non-collinear spin model we expect the same result to hold (up to prefactors) for the spin--spin correlation function. This implies the existence of phase with QLRO in the spin sector both at finite $T$ and at finite $\Delta$, the two being adiabatically connected.


\subsection{Disorder-driven vortex unbinding}

At larger disorder, the analysis of Gaussian fluctuations is insufficient, and vortex physics comes into play. As pointed out in previous literature on the RPXY model, the energy of a single vortex is renormalized by disorder, such that vortices proliferate beyond a critical level of disorder, $\Delta_c$, in a KT-like transition, destroying QLRO.
Early literature on the RPXY model\cite{granato86} suggested that the phase boundary in the temperature--disorder phase diagram displays re-entrant behavior, such that $\Delta_c(T)$ is finite for all temperatures below the clean-limit KT temperature, but displays a maximum at intermediate $T$ and vanishes in the limit $T\to0$.
Subsequent work\cite{nattermann95,cha95,koster97} has, however, argued that re-entrant behavior does not occur, and that $\Delta_c$ remains finite in the limit $T\to0$. The decay exponent $\eta$ at this transition has been proposed to be $\eta_c=1/16$,\cite{nattermann95} to be contrasted with $\eta_c=1/4$ established for the thermal vortex unbinding transition of a standard two-dimensional \emph{XY} model.

However, the details of the disorder-driven vortex unbinding transition are highly non-trivial, as the single-vortex picture is inadequate because it ignores the correlation between the vortex dipole pairs on top of the random dipole background.\cite{tang96} As a result, it presently believed that the critical value of $\Delta$ is non-universal, and the value of the decay exponent is bound between $1/16\leq\eta_c\leq1/4$.\cite{tang96,maucourt97}

For the easy-plane non-collinear magnets of interest, we may expect a similar vortex unbinding transition at $T=0$. However, a key difference to the RPXY model is the presence of chirality order: Chirality order survives small bond disorder as the latter acts like a random-mass term only. Hence, vortex unbinding may occur in the presence of chirality order, with the latter being destroyed at a larger level of disorder, or a novel disorder-driven phase transition with simultaneous destruction of spin QLRO and chirality order may be realized.


\subsection{\emph{XY} vs. Heisenberg}

For the \emph{XY} case we have argued in favor of a $T=0$ spin-QLRO phase at small finite disorder. In contrast, the Heisenberg case displays -- in the same regime -- short-range order only.\cite{dey19}

The reason can be traced to the existence of multiple Goldstone modes in the Heisenberg case, as opposed to a single one in the \emph{XY} case; we recall that in both cases disorder only couples to {\em one} Goldstone mode. The renormalization-group (RG) treatment presented, e.g., in Refs.~\onlinecite{hasselmann04,dey19} shows that the coupling of the disorder-affected Goldstone mode to the other ones leads to a non-trivial RG flow. This makes disorder a marginally relevant perturbation in $d=2$ and eventually generates a mass, implying the destruction of LRO. In contrast, in the \emph{XY} case such non-trivial RG flow is absent. Hence, disorder is exactly marginal, translating into a line of fixed points with QLRO.


\section{Triangular-lattice \emph{XY} antiferromagnet: Numerics}
\label{sec:num}

To verify the above analytical considerations and to provide quantitative results, we have performed large-scale numerical simulations for the triangular-lattice \emph{XY} antiferromagnet with bond disorder. We restrict ourselves to the classical limit and treat the spins $\vec{S}_i$ as unit vectors. For concreteness, we have focused on a bimodal distribution of nearest-neighbor exchange couplings $J_{ij}$, with its width\cite{deltanote} $\Delta$ parametrizing the disorder strength:
\begin{equation}
    \begin{aligned}
        P(J_{ij}) &=
        \begin{cases}
            \frac{1}{2} & \text{for~} J_{ij} = J(1 - \Delta)\\
            \frac{1}{2} & \text{for~} J_{ij} = J(1 + \Delta)
        \end{cases}
    \end{aligned}\label{eq:bimodal_distro}
\end{equation}
However, the qualitative properties will not depend on the precise form of the distribution of
bond randomness.

\subsection{Methods}

Two methods have been employed, namely finite-temperature Monte-Carlo (MC) simulations and zero-temperature energy minimization, both on finite lattices of $N=L^2$ sites with periodic boundary conditions. Our MC moves combine a single-site restricted Metropolis update with microcanonical steps \cite{alonso96} and parallel tempering.\cite{hukushima96} After 10 microcanonical sweeps we performed a Metropolis sweep and a parallel tempering update in sequence. A temperature-dependent selection window pushes the average acceptance rate to larger than 50\% at any temperature. The temperature grid was chosen such that the parallel tempering moves have a success rate larger than 40\%. Such simulation can be efficiently performed down to temperature of $T/J=0.2$.
We compute the disorder-averaged specific heat $c_v$ and the static structure factor
\begin{equation}
S(\vec q) = \frac{1}{L^2} \sum_{ij} [{\langle\vec S_i \cdot \vec S_j\rangle}]_{\rm avg} e^{i\vec q \cdot (\vec R_i-\vec R_j)}
\end{equation}
where $\langle\ldots\rangle$ and $[\ldots]_{\rm avg}$ denote MC and disorder average, respectively. For a state with magnetic LRO at wavevector $\vec Q=(4\pi/3,0)$, the value of $m^2 = \overline{S(\vec Q)}/N$ corresponds to the square of the order parameter $m$ in the thermodynamic limit. Similarly, we determine the chirality structure factor, obtained by replacing ${\vec S}$ by $\kappa$ from Eq.~\eqref{eq:chir} calculated on the upward triangles of the lattice; here order will occur at $\vec Q=0$.
The correlations lengths $\xi^{\rm spin}$ and $\xi^{\rm chir}$ in both the spin and chirality sectors are determined from the corresponding structure factors by extracting $1/\xi$ as the full-width-half-maximum (FWHM) of the peak at $\vec Q$ via a Lorentzian fit.

At zero temperature we employ a separate local energy minimization scheme \cite{henley89} in which, over multiple iterations, we align each spin along the cumulative local magnetic field it experiences. At each iteration all spins are individually reoriented in this manner until the global energy and configurational difference between the iterations fall below a certain threshold. For the \emph{XY} model this scheme suffers from the fact that vortices cannot be efficiently created or destroyed in the course of the iteration. We therefore employ anisotropy annealing, where the simulation starts with $600$ iteration steps using the corresponding Heisenberg model, $\lambda=1$, and then $\lambda$ is continuously reduced to zero over the next $600$ steps of the iteration before the iteration runs until convergence. In addition, we use a large number of initial configurations with varying degree of disorder.


\subsection{Finite-$T$ Monte-Carlo results}

\begin{figure}
\includegraphics[width=0.95\columnwidth]{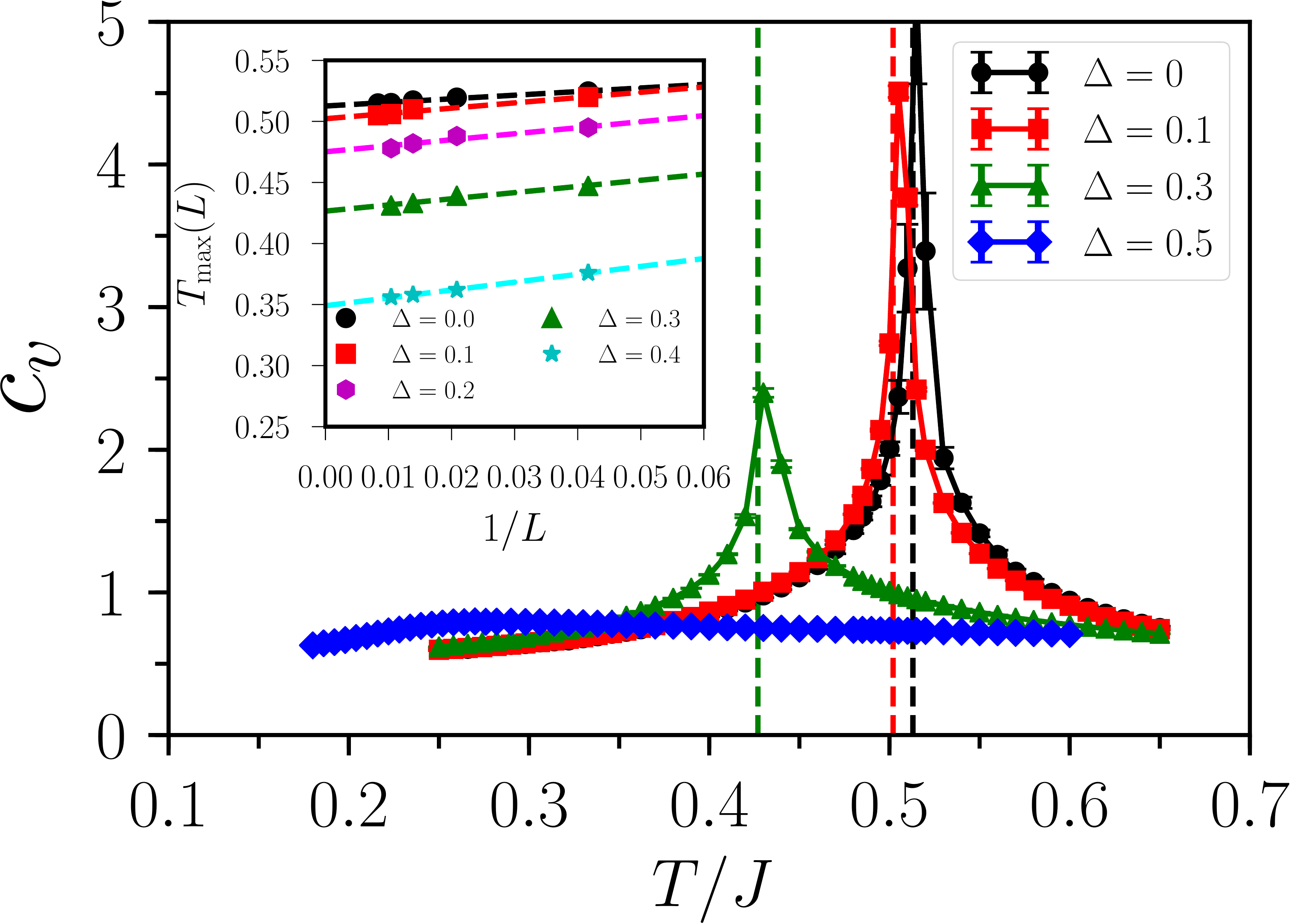}
\caption{
MC results for the specific heat $c_v$ as function of temperature $T$ for the triangular-lattice \emph{XY} model at various value of bond disorder $\Delta$. The main panel shows results for $L=96$, the inset illustrates the finite-size scaling of the peak position in $c_v(T)$. The resulting ordering temperatures are shown as vertical dashed lines in the main plot.
}
\label{fig:cv}
\end{figure}

\begin{figure}
\includegraphics[width=0.513\columnwidth]{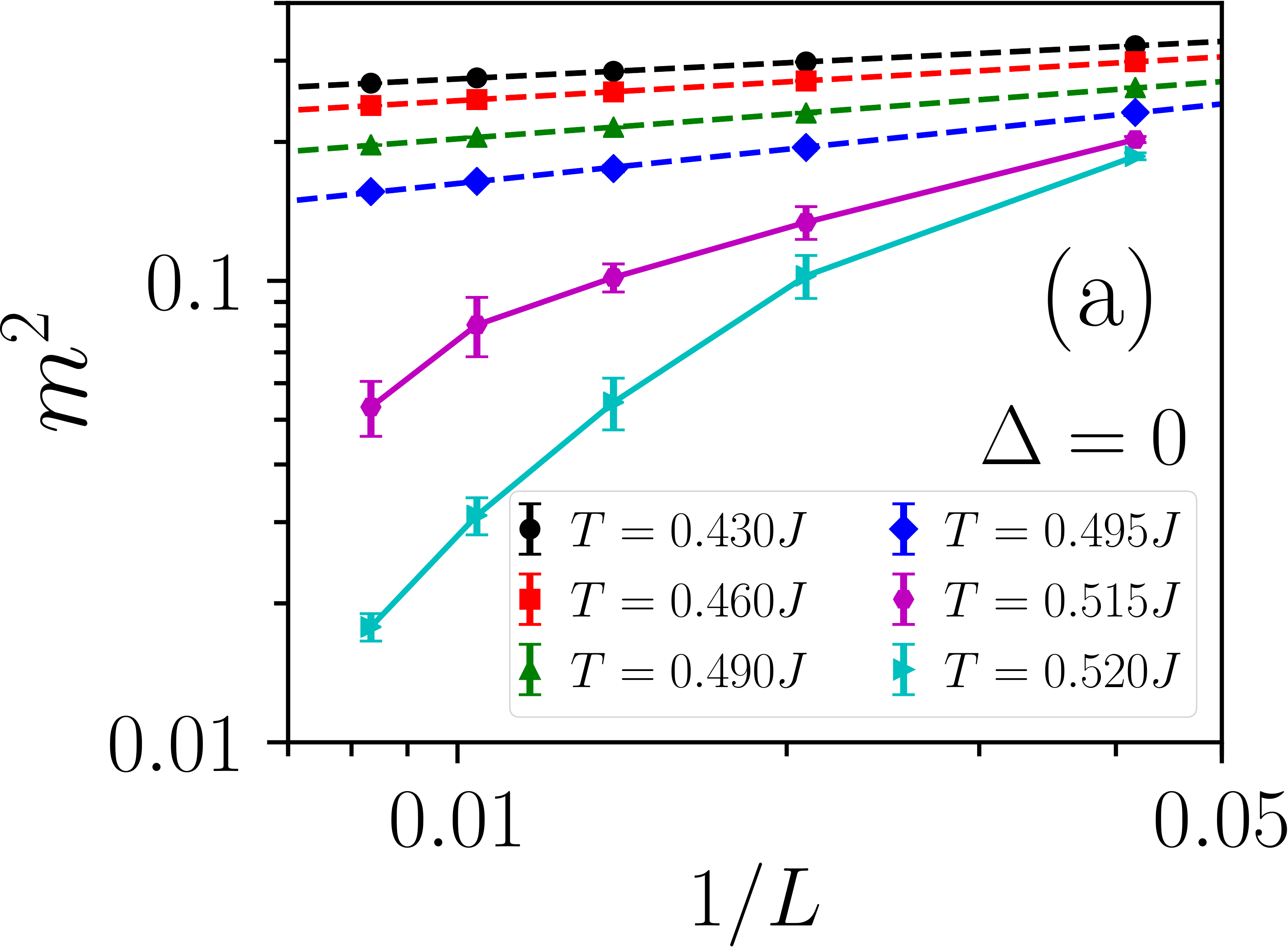}
\includegraphics[width=0.47\columnwidth]{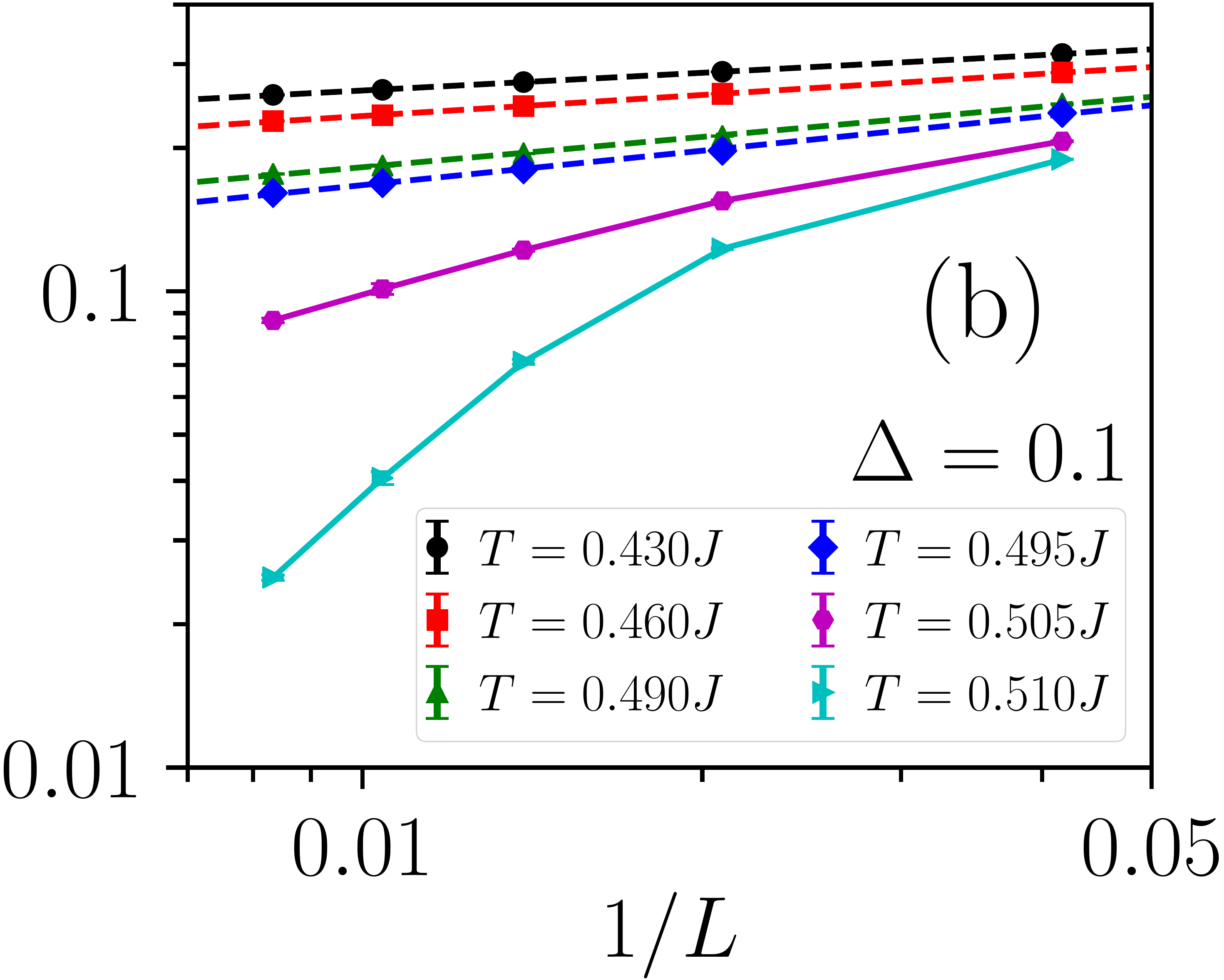}
\includegraphics[width=0.513\columnwidth]{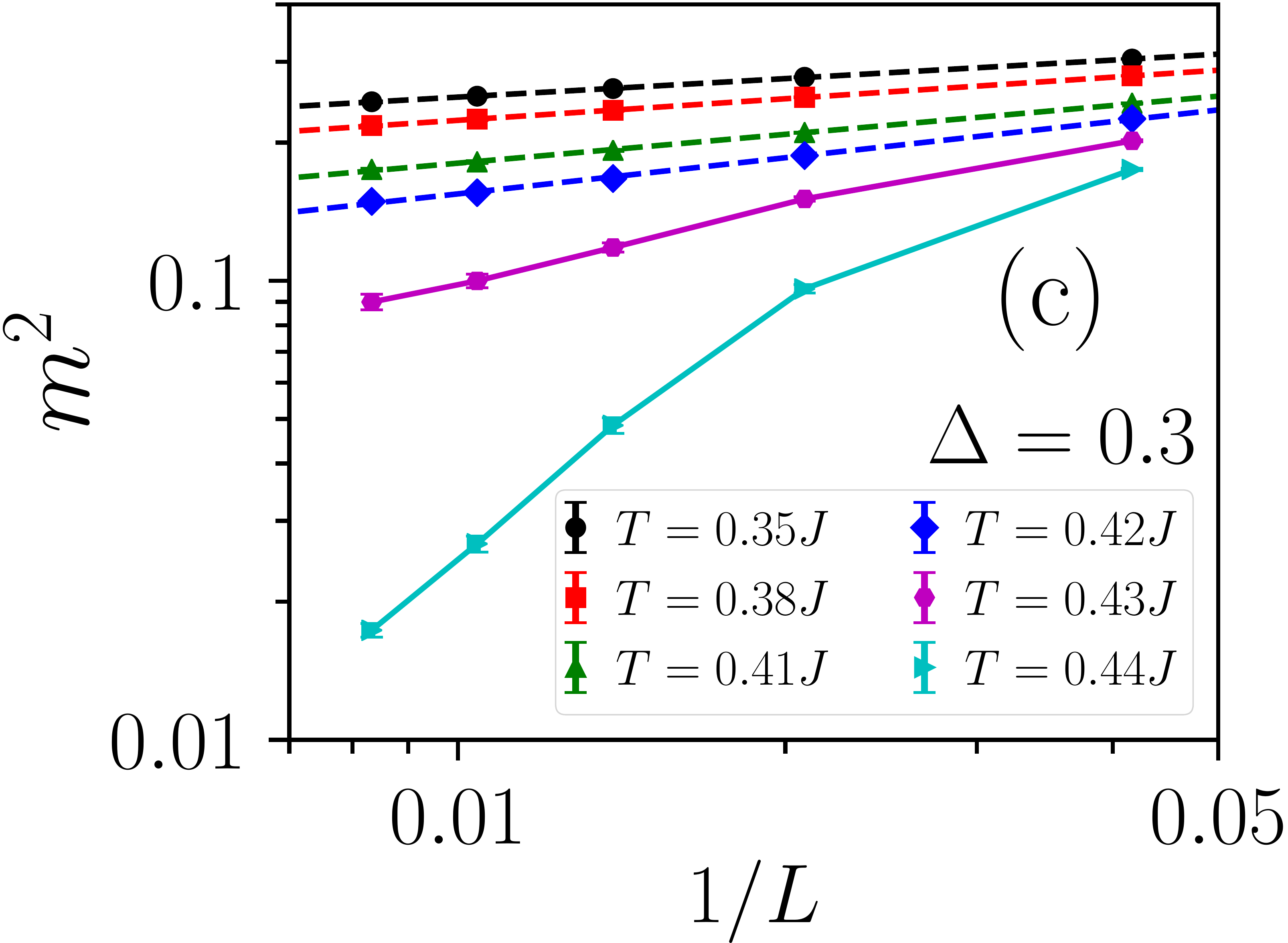}
\includegraphics[width=0.47\columnwidth]{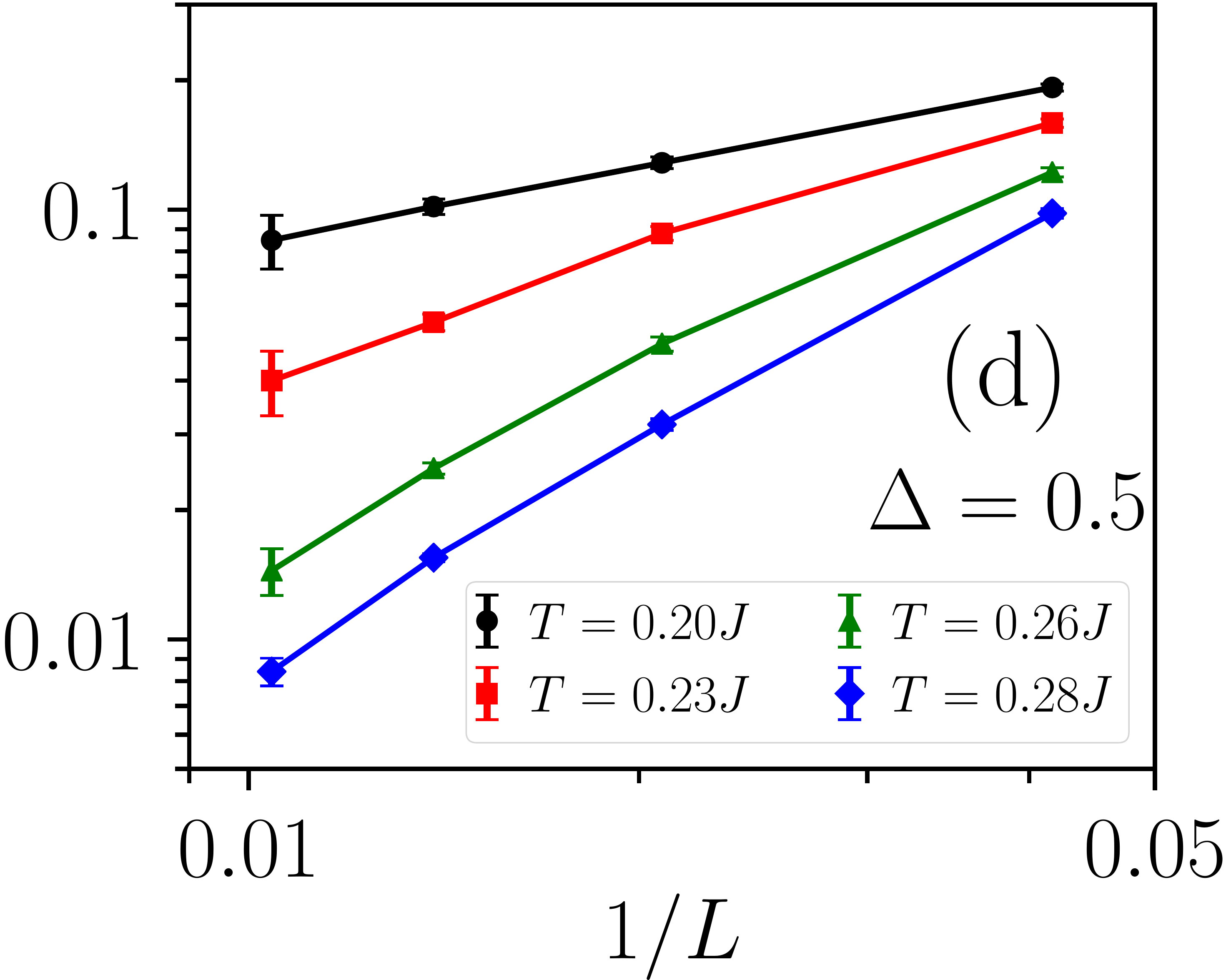}
\caption{
MC results for the size dependence of the squared staggered magnetization $m^2_Q$ for different temperatures $T$ and disorder strengths
(a) $\Delta = 0.0$,
(b) $\Delta=0.1$,
(c) $\Delta=0.3$, and
(d) $\Delta=0.5$.
In (a-c) the low-$T$ data signal a power-law dependence $m^2_Q \propto L^{-\eta}$, as demonstrated by the fit (dashed), whereas the high-$T$ data indicate an exponential $L$ dependence.
}
\label{fig:m2q}
\end{figure}

Sample results from the Monte-Carlo simulations are shown in Figs.~\ref{fig:cv} and \ref{fig:m2q}. For the clean system, we detect a single peak in the specific heat $c_v(T)$, indicating a thermal phase transition.
As mentioned earlier, the model at hand displays a low-temperature chirality order of Ising type, such that two phase transitions -- Ising chirality and KT-like spin -- are expected. As only the Ising transition leaves a strong trace in specific heat, we associate the peak in $c_v(T)$ with the chirality transition. However, extensive numerical work has shown that the two transitions are located extremely close to each other, $\delta T_c/T_c \approx 1/50$, such that only simulations with system sizes $L > 300$ can clearly discern the two transitions.\cite{obuchi12} Our simulations, in particular for the disordered systems, are restricted to $L\leq 96$, therefore both chirality and spin transitions occur together within our resolution.

The specific-heat results in Fig.~\ref{fig:cv} show that weak disorder reduces the transition temperature, while for strong disorder, $\Delta/J>0.5$, no thermal transition can be detected. The peak position in $c_v(T)$ depends weakly on system size, see inset of Fig.~\ref{fig:cv}, and enables to extract the ordering temperature via an extrapolation $L\to\infty$, which according to Ref.~\onlinecite{obuchi12} corresponds to chirality order. In the clean limit, we find $T_c^{\rm chir}/J=0.513(2)$ which agrees with $T_c^{\rm chir}/J=0.51251(1)$ obtained in  Ref.~\onlinecite{obuchi12}.

We separately monitor the finite-size dependence of the staggered magnetization, $m^2(L)$. The clean-limit QLRO phase below the KT-like transition is expected to show a power-law dependence, $m^2\propto L^{-\eta}$, where the exponent $\eta$ increases with increasing temperature and reaches a critical value of $\eta_c=1/4$ at the KT transition.\cite{jknn77} Above the ordering temperature, $m^2(L)$ should decay exponentially.
Our numerical data for the clean system, Fig.~\ref{fig:m2q}(a), is consistent with this, and yield an estimate for the critical temperature of $T_c^{\rm spin}/J = 0.505(5)$, to be compared to $T_c^{\rm spin}/J = 0.5046(10)$ from Ref.~\onlinecite{obuchi12}.

For finite bond disorder, Fig.~\ref{fig:m2q}(b-d), the same analysis can be applied: If we assume $\eta_c=1/4$ for the disordered system then $T_c^{\rm spin}(\Delta)$ can be extracted from the decay $m^2(L)$. $T_c^{\rm chir}(\Delta)$ extracted from the peak location in specific heat is slightly higher, but the two agree within error bars due to our limited resolution.
We recall that $\eta_c$ is expected to decrease with increasing disorder, as noted in Sec.~\ref{sec:rg}, but this effect is likely small up to $\Delta/J=0.45$ where $T_c$ has only decreased by a factor of $2$ compared to the clean system.


\subsection{$T=0$ minimization results}

\begin{figure}
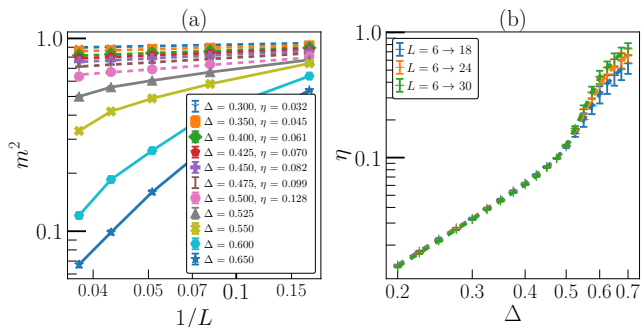

\includegraphics[width=0.49\columnwidth]{{{BIMODAL-SQ-L-XY-JZ-0.00-ALPHA-0.00}}}%
\includegraphics[width=0.49\columnwidth]{{{BIMODAL-ETA-DELTA-XY-JZ-0.00-ALPHA-0.00}}}
\caption
{
(a) $T=0$ minimization results for the size dependence of the squared staggered magnetization $m^2$ for the triangular-lattice \emph{XY} model at different disorder strengths. Also shown is the decay exponent $\eta$ obtained from a power-law fit (shown in dashed lines) 
$m^2\propto L^{-\eta}$ through the data points, $L=6, \ldots, 30$.
(b) Fit exponent $\eta$ as function of disorder strength $\Delta$ and for different $L$ fitting windows.
}
\label{fig:t0_mqeta}
\end{figure}

We now turn to the simulations results obtained directly at $T=0$. As above we monitor the system-size dependence of the staggered magnetization $m^2$. As shown in Fig.~\ref{fig:t0_mqeta}(a) this follows a power law, $m^2\propto L^{-\eta}$, at small disorder. This demonstrates that the 120$^\circ$ LRO present at $\Delta=0$ is replaced by QLRO at finite small $\Delta$, as anticipated above. Moreover, deviations from the power-law behavior are witnessed for $\Delta\gtrsim 0.5$, signaling a $T=0$ transition into a phase with short-range order. Given the classical nature of the spins, this is expected to be a glassy phase \cite{villain79}; we have checked that the corresponding Edwards-Anderson order parameter \cite{fischer} is indeed non-zero.

The exponent $\eta$ obtained from power-law fits for different $\Delta$ is displayed in Fig.~\ref{fig:t0_mqeta}(b). $\eta$ depends quadratically on $\Delta$ for small $\Delta$, as derived in Sec.~\ref{sec:rg}, see Eq.~\eqref{eq:eta}. A clear upward deviation, combined with a significant dependence on the $L$ fitting window, indicates a transition out of the QLRO phase around $\Delta=0.5$.
At $\Delta_c^{\rm spin}\approx0.5$ we read off a critical value of the decay exponent $\eta_c\approx 0.1$. This value is substantially larger than the proposed $\eta_c=1/16$,\cite{nattermann95} supporting the idea of a non-universal $\eta_c$ for the disorder-driven transition.\cite{tang96,maucourt97}

\begin{figure}
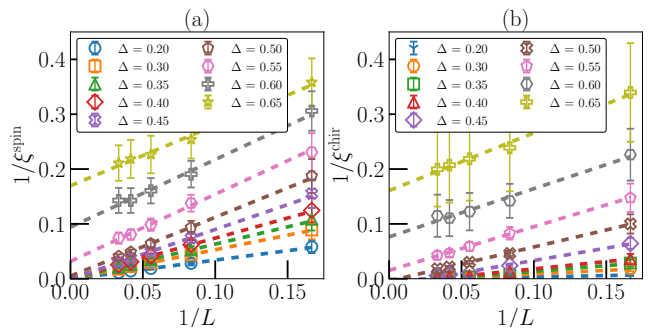

\includegraphics[width=0.49\columnwidth]{{{BIMODAL-SQ-CORRFS-XY-JZ-0.00-ALPHA-0.00}}}%
\includegraphics[width=0.49\columnwidth]{{{BIMODAL-CQ-CORRFS-XY-JZ-0.00-ALPHA-0.00}}}
\caption{
Finite-size scaling of the (a) magnetic and (b) chirality inverse correlation length at $T=0$ for different disorder strengths $\Delta$.
}
\label{fig:t0_corrlen_FS}
\end{figure}

The correlation lengths for both spin and chirality are displayed in Fig.~\ref{fig:t0_corrlen_FS}, which shows the finite-size scaling of $1/\xi$. The data clearly indicate a state with short-range order in both the spin and chirality sectors for large disorder $\Delta$, while signaling infinite correlation length at small $\Delta$. The transition is found around $\Delta_c=0.5$ in both sectors.

To locate the transition more precisely, we analyze the finite-size trend of $\xi/L$ in Fig.~\ref{fig:t0_corrlen}: $\xi/L$ is typically $L$-independent for a scale-invariant critical state, while it increases (decreases) with $L$ for a long-range-ordered (disordered) state, respectively. The data in Fig.~\ref{fig:t0_corrlen}(b) signify a chirality transition at $\Delta_c^{\rm chir}=0.49(1)$ where a clear crossing point of the $\xi(\Delta)/L$ curves is seen. For the spin sector, Fig.~\ref{fig:t0_corrlen}(a) indicates a transition between QLRO - where $\xi/L$ is $L$-independent -- and short-range order which occurs in close proximity,  $\Delta_c^{\rm spin}=0.49(2)$.

\begin{figure}
\includegraphics[width=0.49\columnwidth]{{{BIMODAL-SQ-CORR-XY-JZ-0.00-ALPHA-0.00}}}
\includegraphics[width=0.49\columnwidth]{{{BIMODAL-CQ-CORR-XY-JZ-0.00-ALPHA-0.00}}}
\caption{
(a) Magnetic and (b) chiral correlation length at $T=0$, plotted as $\xi/L$ as function of disorder strength $\Delta$. Vanishing $L$ dependence signals a scale-invariant state, and crossing point therefore indicates a second-order transition.
The data in (a) demonstrate a transition from QLRO to short-range order in the spin sector, while (b) exhibits an order-disorder transition in the chirality sector, with the vertical dashed line marking the crossing point of $\xi/L$.
}
\label{fig:t0_corrlen}
\end{figure}


\begin{figure}
\includegraphics[width=0.9\columnwidth]{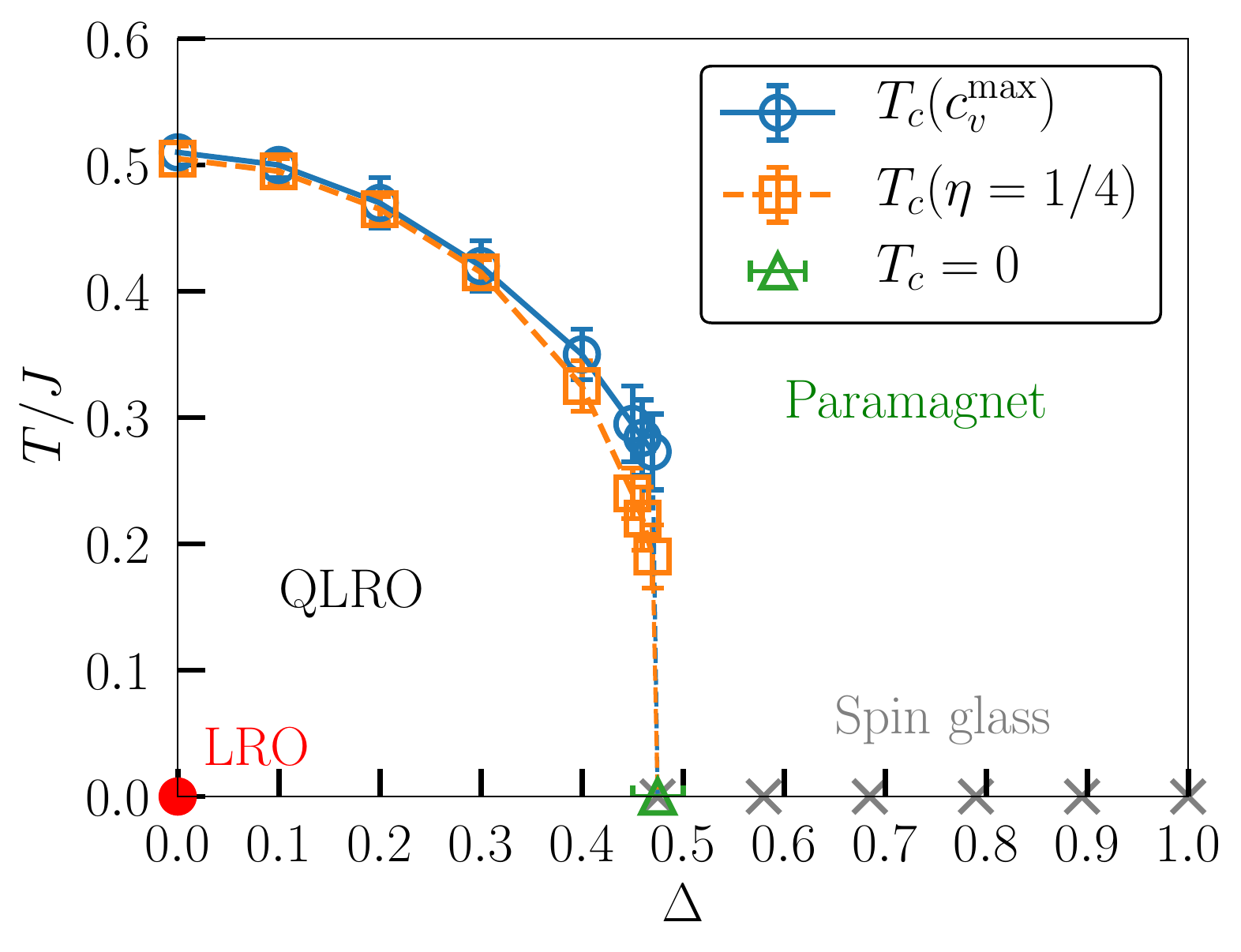}
\caption{
Phase diagram for the triangular-lattice \emph{XY} model with bimodal bond disorder, shown as function of temperature $T$ and disorder strength $\Delta$, as constructed from the results of finite-$T$ MC simulations and $T=0$ minimization.
We have extracted the chirality transition, $T_c(c_v^{\rm max})$, and the spin transition, $T_c(\eta=1/4)$, via separate criteria. However, clearly resolving an intermediate phase is beyond our numerical resolution, for details see text.
}
\label{fig:xy_pd}
\end{figure}

\subsection{Phase diagram}

We are now in the position to construct the phase diagram of the bond-disordered triangular-lattice \emph{XY} model as function of temperature $T$ and disorder strength $\Delta$, this is shown in Fig.~\ref{fig:xy_pd}.
120$^\circ$ LRO is realized at $T=0$ and $\Delta=0$; it gives way to QLRO for any small finite $T$ and/or $\Delta$. We recall that bond disorder does not couple to chirality, i.e., does not act like a random field, hence the QLRO phase display chirality LRO also at finite $\Delta$, consistent with Fig.~\ref{fig:t0_corrlen}(b). At large $\Delta$ QLRO is destroyed at $T=0$ in favor of a glassy phase. Given that $d=2$ is below the lower critical dimension for a finite-temperature spin-glass transition \cite{parisi18}, there is no glass transition at finite $T$, but instead the small-$T$ large-$\Delta$ regime is adiabatically connected to the high-$T$ paramagnet.

As discussed above, going from QLRO to the paramagnetic phase likely involves a sequence of two nearby transitions, associated with spin and chirality degrees of freedom, respectively. However, our numerics can only resolve one transition due to limited system size.
The corresponding finite-$T$ phase boundary connects to the $T=0$ transition between QLRO and spin glass. The shape of the phase diagram then indicates non-reentrant behavior as in the unfrustrated RPXY model.


\section{Conclusions and outlook}

For two-dimensional chiral easy-plane antiferromagnets, we have shown that infinitesimal random-bond disorder destroys the non-collinearly ordered ground state of the clean limit in favor of a QLRO state which is chirally ordered. For stronger disorder this gives way to a short-range-ordered glassy state. The $T=0$ transition between QLRO and spin glass continuously connects to its finite-temperature counterpart, the QLRO--paramagnet transition, see Fig.~\ref{fig:xy_pd}. Within our numerical accuracy we are not able to resolve whether this transition happens in a single step or two steps, i.e., via an intermediate state with chiral long-range order but spin disorder.
Our results show that the behavior of the frustrated random-bond \emph{XY} antiferromagnet is qualitatively different from that of a fully frustrated random-phase \emph{XY} model. In the latter, any order is destroyed already by infinitesimal disorder by a random-field effect \cite{gupta99}.

We have also performed simulations for the triangular-lattice \emph{XXZ} model \eqref{eq:xxz} with anisotropy $0<\lambda<1$, in order to connect the present results to those for the Heisenberg case.\cite{syrom15,dey19} Remarkably, for all $\lambda<0.9$ we find results (not shown) which are quantitatively very similar to that of the \emph{XY} case, $\lambda=0$. Minor changes can be detected for $\lambda=0.95$, but our numerical resolution -- limited by system size -- is not sufficient to fully map out the phase boundaries for $\lambda\to1^-$. We leave this task for future work, which also involves to clarify the fate of the {\Ztwo} vortex transition in the Heisenberg case \cite{kawamura10} upon lowering $\lambda$.

A number of layered non-collinear easy-plane magnets on the triangular lattice have been reported, e.g., RbFe(MoO$_4$)$_2$ (Refs.~\onlinecite{svistov03,white13}), Ba$_3$CoSb$_2$O$_9$ (Refs.~\onlinecite{doi04,susuki13}), KAg$_2$Fe[VO$_4$]$_2$ and RbAg$_2$Fe[VO$_4$]$_2$ (Ref.~\onlinecite{moeller14}), and RbFeCl$_3$ (Ref.~\onlinecite{wada82}). Our \emph{XY} model results are of direct relevance for these materials in the presence of bond disorder, for instance to Cs$_{1-x}$Rb$_x$FeCl$_3$ studied recently in Ref.~\onlinecite{zheludev19}.

The low-energy physics studied here is also relevant for frustrated Heisenberg magnets in applied magnetic field: The field leaves a {\Uone} symmetry intact, and non-collinear order can develop \emph{perpendicular} to the field, such that the finite-field order is non-coplanar. We note that this does not directly apply to the plain triangular-lattice Heisenberg model, as this realizes ordered states which are coplanar with the field direction due to an order-from-disorder mechanism, but it applies to the same model supplemented with a sufficiently strong biquadratic interaction which stabilizes non-coplanar umbrella-type states.
For such magnets, the underlying Heisenberg symmetry implies that any amount of random-bond disorder destroys $T=0$ LRO in favor of a spin glass.\cite{dey19} An applied field renders the spontaneous order \emph{XY}-like, and a QLRO phase appears at finite small $\Delta$. We hence expect that the transition between QLRO and glass cannot only be driven by varying the disorder level, but also the applied field. A detailed investigation of the resulting transitions is left for future work.


\begin{acknowledgments}
We are grateful to M. Gingras, J. A. Hoyos, T. Vojta, A. Zheludev, and M. Zhitomirsky
for instructive discussions.
SD and MV acknowledge financial support from the Deutsche Forschungsgemeinschaft (DFG) through SFB 1143 (project-id 247310070) and the W\"urzburg-Dresden Cluster of Excellence on Complexity and Topology in Quantum Matter -- \textit{ct.qmat} (EXC 2147, project-id 390858490). ECA was supported by CNPq (Brazil) Grants No. 406399/2018-2 and 302994/2019-0, and FAPESP (Brazil) Grant No. 2019/17026-9.
\end{acknowledgments}



\begin{thebibliography}{99}
\bibitem{deltanote}
In our numerics, $\Delta$ characterizes the strength of the bimodal disorder; this is similar but not identical to the $\Delta$ appearing in Eq.~\eqref{eq:eta} due to different forms of quenched disorder.
\end{thebibliography}
\end{document}